\documentclass[twocolumn,showpacs,preprintnumbers,amsmath,amssymb]{revtex4}

\usepackage{graphicx}
\usepackage{dcolumn}
\usepackage{bm}

\begin{document}

\preprint{}

\title{Tunable quantum interference between noisy electron sources}

\author{Yuanzhen Chen}
\author{Samir Garzon}
\author{Richard A. Webb}%

\affiliation{Department of Physics and USC NanoCenter, University
of South Carolina, Columbia, South Carolina 29208, USA}

\date{\today}

\begin{abstract}
We report shot noise cross correlation measurements in a four
terminal beam splitter configuration. By using two tunnel barriers
as independent electron sources with tunable statistics and
energy, we can adjust the degree of quantum interference that
results when the electrons scatter at a beam splitter.
Even though quantum interference is only weakly affected by noise,
it can be strongly suppressed by detuning the energies of the
interfering electrons. Our results illustrate the importance of
indistinguishability for quantum interference, and its resilience
to unsynchronized electron sources and noise.

\end{abstract}

\maketitle


Shot noise measurements~\cite{Blanter2000:PR} in multi-terminal
mesoscopic devices have been used to probe quantum statistics and
interference
\cite{Loudon1998:PRA,Liu1998:nature,Oliver1999:science,
Henny1999:science,Vishveshwara2003:PRL}, to understand the
dynamics of electrons and their interaction
\cite{Samuelsson2002:PRL,Borlin2002:PRL,
Cottet2004:PRB,Bignon2004:EPL,Oberholzer2006:PRL,Zhang2007:PRL},
and have even been proposed as probes for electron quantum
entanglement
\cite{Loss2000:PRL,Samuelsson2003:PRL,Chtchelkatchev2002:PRB,Egues2002:PRL,
Lebedev2005:PRB,Burkard2003:PRL,Lebedev2004:PRB}, which has not
yet been observed. To date, most experimental reports on electron
quantum statistics and interference use three terminal
configurations with a single electron
source~\cite{Oliver1999:science,
Henny1999:science,Oberholzer2006:PRL,Zhang2007:PRL}, even though
configurations which use two electron sources are essential for
measuring electron quantum entanglement. In the only two-electron
source measurement reported to date, Liu \textit{et
al.}~\cite{Liu1998:nature} used shot noise cross correlation
measurements to observe destructive quantum interference between
two noiseless electron sources. However, it was pointed out that
in order to proceed towards the observation of electron
entanglement, the effects of noise and lack of synchronization on
quantum interference had to be understood~\cite{Hu2004:PRB}. In
this Letter we use shot noise cross correlation measurements to
observe quantum interference between two noisy electron sources.
These tunnel barrier electron sources are unique since (i) the
statistics of the tunneling currents can be tuned over a wide
range and (ii) the energy of the tunneling electrons can be
precisely controlled. Our data indicates that quantum interference
between electrons from the two uncorrelated and unsynchronized
sources can occur even in the presence of noise. Even though
quantum interference is only weakly dependent on the amount of
noise present in the sources, it can be strongly suppressed by
decreasing the energy overlap between electrons from both sources.
Our experiments thus provide a direct observation of the
fundamental relation between indistinguishability and quantum
interference.

\begin{figure}[!tb]
\includegraphics[width=8.5cm,angle=0]{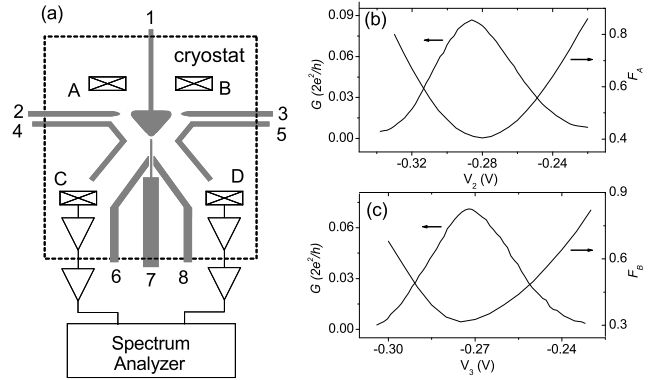}
\caption{\label{fig1} (a) Device schematic. $A$, $B$, $C$, and $D$
are electron reservoirs, while gates 1, 2, and 3 define the two
tunnel barriers. Large negative voltages are applied to gates 6
and 8 so that the partition of electrons occurs only at the narrow
section of gate 7 between the bottom of gate 1 and the top of
gates 6 and 8. (b) \& (c) Conductance and Fano factor of the (b)
left and (c) right tunnel barriers.}
\end{figure}

The samples used in this experiment were defined on a GaAs/AlGaAs
heterostructure using electrostatic gates~\cite{Chen2006:PRB}. At
low temperatures a two dimensional electron gas (2DEG) with
electron mobility $\mu_e=6.1\times10^5$ cm$^2$/Vs and carrier
density $n=1.8\times10^{15}$ m$^{-2}$ forms 50 nm below the wafer
surface. A schematic of the gates is shown in Fig. \ref{fig1}(a).
Negative voltages are applied to gates 1, 2 and 3 to form two
tunnel barriers, and to gates 6, 7, and 8, to form a beam splitter
at the thin section of gate 7. The transmission coefficient $t$ of
the beam splitter can be adjusted by changing the voltage on gate
7. Electrons injected from reservoirs $A$ and $B$ tunnel through
the barriers and are guided by additional gates 4 and 5 towards
the beam splitter, where they scatter into channels $C$ and $D$.
Mean free paths in our devices are $\sim$1$\mu$m and thus
electrons travel ballistically from the tunnel barrier to the beam
splitter. The current fluctuations in both channels are measured
by two cryogenic preamplifiers, further amplified at room
temperature, and eventually fed into a spectrum analyzer, which
calculates their cross correlation $S$~\cite{Blanter2000:PR}. All
measurements are done in a 20 kHz window around 220 kHz and at a
temperature of 70 mK. Details of the measurement setup and of the
thermal noise background subtraction procedure used to obtain the
current shot noise are described in detail
elsewhere~\cite{Chen2006:PRB}. We studied a total of 4 devices, repeating the measurements on each device after multiple room temperature thermal cycles. The same general behavior was observed every time. Except when explicitly noted, the data reported here is
for a single device on a single cooldown.

\begin{figure}[!tb]
\includegraphics[width=8.5cm,angle=0]{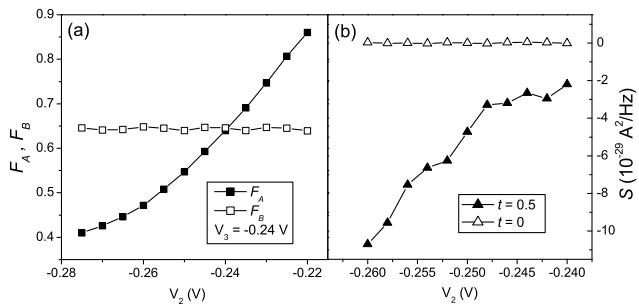}
\caption{\label{fig2} (a) $F_{A}$ and $F_{B}$ as functions of
$V_{2}$ showing that the statistics of the two sources can be
tuned independently. (b) $S$ as a function of $V_{2}$ at $t$ = 0.5
and $t$ = 0 showing that the two sources are uncorrelated.}
\end{figure}

Tunnel barriers fabricated in semiconductor heterostructures
usually contain localized electronic states
\cite{Chen1992:PRB,Safanov2003:PRL,Kuznetsov2000:PRL} which can be
probed with conductivity and shot noise measurements. It has been
previously reported that by tuning the gate voltages used to form
such tunnel barriers, the energy of the localized states and the
coupling between them and the electron reservoirs can be
adjusted~\cite{Safanov2003:PRL}. This modifies not only average
transport quantities such as the tunneling current $I$, but also
fluctuations dependent on the electron transport statistics, such
as the shot noise. Therefore a tunnel barrier is an electron
source with tunable statistics characterized by the Fano factor
$F$~\cite{Blanter2000:PR}, defined as the ratio of the total shot
noise current power and 2$eI$.

We first characterize each of the tunnel barriers by measuring
their conductivity and Fano factor as a function of gate voltage
[Figs. 1(b) and 1(c)]. Shot noise suppression below the ideal
tunnel barrier value of $F=1$ occurs when transport is dominated
by conduction through localized states, as also evidenced by the
strong conductance modulation with gate voltage. Therefore by
adjusting $V_2$ ($V_3$) we can control the statistics of electrons
coming from the left (right) tunnel barrier. Furthermore, we
observe that the two electron sources can be tuned independently.
Figure \ref{fig2}(a) shows that as $V_{2}$ changes, the Fano
factor of the left barrier is strongly modulated, while the Fano
factor of the right barrier remains constant. Similarly (data not
shown), the properties of the left barrier are unmodified when
$V_3$ is changed. In order to show that the two tunnel barriers
inject uncorrelated electrons, we measured the cross correlation
$S$ of the signals at electron reservoirs $C$ and $D$ for both
$t$=0 (beam splitter completely closed) and $t$ = 0.5. Figure
\ref{fig2}(b) shows that there is zero cross correlation when the
beam splitter is completely closed, but a clear gate voltage
dependent cross correlation when $t$ = 0.5, demonstrating that the
only significant source of correlations occurs when the electrons
scatter at the beam splitter. Therefore, we conclude that the
tunnel barriers act as uncorrelated sources of electrons with
independently adjustable statistics.

\begin{figure}[!tb]
\includegraphics[width=8.5cm,angle=0]{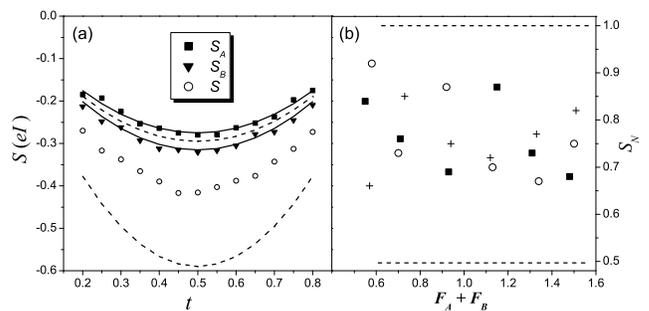}
\caption{\label{fig3} (a) Single source cross correlations
($S_{A}$ and $S_{B}$) and dual source cross correlation ($S$) as
functions of $t$ for $F_{A}$ = 0.45 and $F_{B}$ = 0.37. Solid
curves are predictions using Eq. 1 and the measured values of
$F_i$, $I_i$, and $t$. Lower dashed curve is the sum of the two
solid curves while upper dashed curve is one half of that sum. (b)
$S_{N}$ at $t$ = 0.5 as a function of $F_{A} + F_{B}$ for three
different samples (for these measurements, $I_{A} = I_{B} = I$). Fluctuating patterns
for each sample are reproducible in repetitive measurements (the
size of typical error bars is $\pm$0.01).}
\end{figure}

Our goal is to quantify the quantum interference between these two
electron sources by measuring the cross correlation at reservoirs
$C$ and $D$. However, even single source injection produces a
nonzero cross correlation~\cite{Oliver1999:science,
Henny1999:science,Oberholzer2006:PRL,Zhang2007:PRL,Chen2006:PRL}.
Therefore, we first perform single source cross correlation
measurements. Electrons are injected from reservoir $A$($B$)
through a single tunnel barrier while the shot noise cross
correlation $S_{A}$($S_{B}$) between reservoirs $C$ and $D$ is
measured. The subindexes $A$ and $B$ indicate which tunnel barrier
was used as an electron source. It was previously
found~\cite{Chen2006:PRL} that $S_{A}$($S_{B}$) is related to
$F_{A}$($F_{B}$), the Fano factor of the source barrier, by

\begin{equation}
\label{equ:xspec2} S_{i} = 2eI_{i}(F_{i}-1)t(1-t),
\end{equation}

\noindent where $i=A,B$. The solid symbols in Fig. \ref{fig3}(a)
show typical single source cross correlation measurements as a
function of the beam splitter transmission coefficient. For the
data shown here $F_A$ = 0.45 and $F_B$ = 0.37, but good agreement
of the single source cross correlation with Eq.~\ref{equ:xspec2}
(solid curves) was found for all other measured values of the Fano
factors.

For the dual source experiments, electrons are injected from both
reservoirs while the shot noise cross correlation $S$ between
reservoirs $C$ and $D$ is measured. The dual source cross
correlation data [open circles in Fig.~\ref{fig3}(a)] show a
similar dependence on $t$, but with larger negative values. For
uncorrelated sources which inject distinguishable particles,
electrons from different sources scatter at the beam splitter
independently, so the total cross correlation is simply
$S_{indep.}=S_{A}+S_{B}$ [lower dashed line in
Fig.~\ref{fig3}(a)]. On the other hand, for uncorrelated sources
of identical particles, quantum interference due to electron
wavefunction overlap at the beam splitter needs to be taken into
account. Theory predicts that for two uncorrelated, noiseless
electron sources ($F_A=F_B=0$), $S_{ideal} = -2eIt(1-t)$
\cite{Buttiker1992:PRB}, where $I$ is the average current in each
channel, and thus $S_{ideal}$ is only one half of $S_{indep.}$
[upper dashed line in Fig.~\ref{fig3}(a)]. Since for every $t$ the
measured dual source cross correlation is less than that for the
case of uncorrelated and distinguishable particles
($|S|<|S_{indep.}|$), we can conclude that there is quantum
interference between electrons arriving from the two different
sources as they scatter at the beam splitter. It was pointed out
that quantum interference can only occur when there is
simultaneous arrival of electrons from both sources at the beam
splitter, or equivalently, that there should be enough
wavefunction overlap between pairs of electrons at the beam
splitter to make the particles
indistinguishable~\cite{Hu2004:PRB}. However, this is not
guaranteed if noisy sources are used, since in that case the time
interval between successive electron arrivals at the beam splitter
is a random variable, and thus interfering electrons are unsynchronized. As such,
electrons from two uncorrelated and noisy sources could have
little or no spatial overlap and quantum interference might not
occur. Nevertheless, the data of Fig.~\ref{fig3}(a) clearly show
that such quantum interference does exist for noisy sources.


To better characterize the effect of quantum interference, we now
define a dimensionless quantity $S_{N}$ = $S/(S_{A}+S_{B})$.
Explicitly,
\begin{equation}
\label{equ:S} S_{N} =
S/((2eI_{A}(F_{A}-1)+2eI_{B}(F_{B}-1))t(1-t)).
\end{equation}
\noindent With this definition, uncorrelated and distinguishable
particles, which present no quantum interference, have $S_N$=1. On
the other hand, for uncorrelated and noiseless sources emitting
indistinguishable particles, quantum interference is maximum and
$S_{N}$=0.5. The results of measurements of $S_N$ as a function of
$F_A+F_B$ for $t$=0.5 for three different samples are shown in
Fig.~\ref{fig3}(b). For each of the samples and each of the $F_A$,
$F_B$ combinations we first obtained data similar to that shown in
Fig.~\ref{fig3}(a). Good agreement of $S_A$ and $S_B$ with Eq.
(\ref{equ:xspec2}) was always observed. Figure~\ref{fig3}(b) shows
that for all $F_{A}$, $F_{B}$ combinations, there is always some
degree of quantum interference ($S_N<1$), but ideal quantum
interference is never obtained ($S_N>0.5$ even for the sources
with the smallest Fano factors). In addition, we do not observe
any clear relation between $S_{N}$ and $F_{A}$, $F_{B}$. This is in
contrast with the results of single source cross correlation
measurements where $S_{i}$ is determined by $F_{i}$ as shown by
Eq. (\ref{equ:xspec2}). As we will now show, the degree
of quantum interference measured by the dimensionless cross
correlation $S_N$ is determined mainly by the energy overlap of the
electrons coming from the two sources, and not by the tunnel barrier Fano factors.

\begin{figure}[!tb]
\includegraphics[width=8.5cm,angle=0]{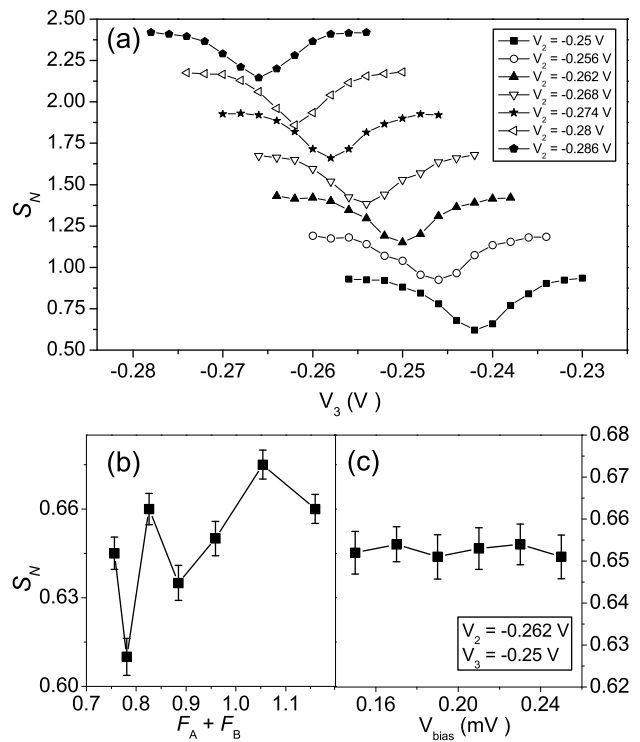}
\caption{\label{fig4} a) $S_{N}$ at $t$ = 0.5 as a function of
gate voltage $V_{3}$ at different values of gate voltage $V_{2}$.
Curves are offset vertically by 0.25
units for clarity. (b)
Minimum values of the seven curves in (a) plotted as a function of
$F_{A}+F_{B}$. (c) $S_{N}$ ($t$ = 0.5) as a function of the bias
voltage applied across source barriers at fixed gate voltages
$V_{2}$ and $V_{3}$.}
\end{figure}

Figure \ref{fig4}(a) shows $S_{N}$ as a function of gate voltage
$V_{3}$ at seven different values of gate voltage $V_{2}$. By varying $V_{2}$ and $V_{3}$ the energy of the localized states
through which electrons tunnel in the two source barriers are
changed. In this
set of measurements, $S$ is measured at each $V_{2}, V_{3}$
combination and $S_{N}$ is calculated using Eq. (\ref{equ:S}).
Again, in all measurements $S_{N}$ is always between 0.5 and 1,
suggesting partial quantum interference of electrons. Furthermore,
there is a strong dependence of $S_{N}$ on $V_{2}$ and $V_{3}$.
For every value of $V_{2}$, $S_{N}$ always has a minimum at a
certain value $V_{3}=V_{3,min}$ and approaches 1 as $V_{3}$ is
tuned away from $V_{3,min}$. As $V_{2}$ changes from -0.25V to
-0.286V, the position of the minima shifts linearly from -0.242V
to -0.266V, giving a ratio of $\Delta V_{2}$/$\Delta V_{3}$ = 1.5. The sensitivity of the localized state energy to changes in gate voltages can be independently measured by calculating the full width at half maximum of the conductance curves in Figs. \ref{fig1}(b), (c), which are 52 mV and 34 mV respectively. The ratio $\Delta V_{2}$/$\Delta V_{3}$ obtained in this way is 1.53, in agreement with the ratio obtained from the cross correlation measurements,
suggesting that the gate voltage dependence of $S_{N}$ in Fig.
\ref{fig4}(a) is a measure of the degree of alignment of the
energies of the localized states in the two tunnel barriers.

The importance of the energy overlap between electrons from the
two sources for quantum interference is now explained. In most
theoretical studies of quantum interference in a beam splitter
configuration, electrons are assumed to be in plane wave states
with well defined wave vectors and energy. However, a more
realistic picture is to view electrons as energy wave packets with
a finite energy broadening defined by the coupling between the
localized states and the reservoirs \cite{Hu2004:PRB}. If
electrons from the two sources have similar energies, that is, the
two wave packets have a significant overlap on the energy scale,
then these electrons become indistinguishable when their
wavefunctions overlap at the beam splitter. In such a case,
quantum interference occurs and the shot noise cross correlation
is suppressed [minima in Fig.~\ref{fig4}(a)]. On the other hand if
electrons from the two sources have very different energies, then
we could in principle identify each electron by measuring its
energy (since the distance from the source to the beam splitter is
smaller than the mean free path, electron motion is ballistic and
thus electron scattering can be ignored). In such a case, these
electrons become effectively distinguishable even when their
spatial wavefunctions overlap. As a result, no quantum
interference occurs, and thus there is no shot noise cross
correlation suppression [saturation towards $S_N$=1 in
Fig.~\ref{fig4}(a)].

Figure~\ref{fig4}(b) shows the minimum values (maximum quantum
interference) of the seven curves shown in Fig.~\ref{fig4}(a) as a
function of $F_{A}+F_{B}$. We want to point out that the seemingly
random point to point fluctuations are actually reproducible in
repetitive measurements. The size of the error bars ($\sim$ 0.01)
is only one third of the typical point to point fluctuation. Once
the gate and bias voltages are fixed, the same $S_{N,min}$ values
will be obtained. Our data indicates that the changes in
$S_{N,min}$ with $F_{A}+F_{B}$ are much smaller than the changes
in $S_N$ obtained when the energies of electrons from both sources
change from aligned to not aligned [Fig. \ref{fig4}(a)]. This
shows that even as the Fano factor, and therefore the noise, of
the sources varies by over a factor of two, there is very small
change in the degree of quantum interference. This evidences that
quantum interference is extremely resilient to noise and lack of
synchronization.

We also observed that varying the bias voltage by almost a factor
of two (for the sample studied here between 0.15 mV and 0.25 mV),
has no effect on quantum interference. Figure~\ref{fig4}(c) shows
that as a function of the bias voltage (equal for both barriers),
and for fixed gate voltages $V_{2}$ and $V_{3}$, $S_{N}$ varies by
less than 1$\%$, a few times smaller than the variation of $S_{N}$ with F shown
in Fig. \ref{fig4}(b). As long as the bias voltage is varied over
this range, tunneling should occur through the same set of
localized states in each barrier. Since $V_{2}$ and $V_{3}$ are
fixed, the energy of these states remains unchanged, so varying
the bias voltage should have negligible effect on tunneling, and
thus no effect on the quantum interference occurring at the beam
splitter. Therefore quantum interference is very insensitive to
the bias voltage, weakly dependent on the source noise, but
strongly affected by the energy overlap of the electrons from each
source.

In summary, we performed shot noise cross correlation measurements
in a four terminal beam splitter configuration using two
uncorrelated tunnel barriers to inject electrons. The observed
shot noise suppression of electrons leaving
the beam splitter is a direct manifestation of quantum
interference. We observe that quantum interference occurs even for
noisy sources and that the degree of quantum interference is only
weakly sensitive to the amount of noise present in the electron
sources. Therefore, our observations show that the synchronization
of electrons is not critical for observing quantum interference.
However, quantum interference can be greatly suppressed by
detuning the energy of the two localized states, since in such a
case electrons from the two sources have different energy and
become distinguishable. Therefore, in order to observe maximum
electron quantum entanglement, it is not the statistical
properties of the electron sources what matters most, but rather
the indistinguishability of the electrons brought about by
carefully aligning their energies.

We want to thank Xuedong Hu for very helpful and stimulating
discussions. This work was supported by NSF through grant No.
DMR0439137.

\bibliography{references}

\end{document}